# An Automatic Volume Control For Preserving Intelligibility

Franklin Felber
Starmark Technologies Division
Starmark, Inc.
P. O. Box 270710, San Diego, CA 92198, USA
Felber@StarmarkTechnologies.com

*Abstract*—A new method has been developed to adjust volume automatically on all audio devices equipped with at least one microphone, including mobile phones, personal media players, headsets, and car radios, that might be used in noisy environments, such as crowds, cars, and outdoors. The method uses a patented set of algorithms, implemented on the chips in such devices, to preserve constant intelligibility of speech in noisy environments, rather than constant signal-to-noise ratio. The algorithms analyze the noise background in real time and compensate only for fluctuating noise in the frequency domain and the time domain that interferes with intelligibility of speech. Advantages of this method of controlling volume include: Controlling volume without sacrificing clarity; adjusting only for persistent speech-interference noise; smoothing volume fluctuations; and eliminating static-like bursts caused by noise spikes. Practical human-factors approaches to implementing these algorithms in mobile phones are discussed.

I. INTRODUCTION AND BACKGROUND

This paper presents a method for automatically adjusting the volume of an audio device to compensate only for noise that interferes with the intelligibility of speech or appreciation of music from the audio device.

The automatic volume control (AVC) described here [1] is a fully automatic system and method for adjusting the volume of an audio output device, such as a mobile phone or car radio, in accordance with listener preferences, to compensate selectively for changing levels of ambient noise only in the time and frequency domains that interfere with intelligibility of speech or appreciation of music.

An example of an audio device is a car radio. Many sources of noise can interfere with hearing a car radio, including tire (road) noise, wind, engine noise, traffic (highway) noise, the fan of a heater or air conditioner, and noises made by the driver and passengers. The noise levels of all of these sources can change with time, depending on factors like the speed of the car or changing environmental conditions outside or inside the car. The noise levels can change abruptly or quasi-continuously or can be transient. Having repeatedly to manually adjust the volume of an audio device to compensate for changing noise levels is a nuisance, and, in a car, can compromise the safety of the occupants and others.

Not all noise, however, interferes with a listener's understanding or appreciation of the output of an audio device. And not all noise, therefore, would impel a listener to want to change the volume. For example, nearly all the information in speech is contained within the frequency interval 200 Hz to 6 kHz [2]. Generally, only the frequency components of noise within this interval can detract significantly from intelligibility of speech. Similarly, the intelligibility of full sentences in noisy environments is substantially greater than the intelligibility of isolated words. Generally, only noises that persist long enough to mask more than a few words can detract significantly from intelligibility of speech [2].

Any system that attempts to compensate for all noise, regardless of frequency or duration, will generally overcompensate by raising or lowering the volume of an audio device to adjust for noise that is not significantly interfering with the ability to listen to the audio device. For example, the occurrence of a high-pitched whine above 6 kHz should not generally be cause for the volume of an audio device to be increased automatically, or to be decreased upon its cessation. Similarly, a transient noise within a car, or another car passing at high speed in the opposite direction, should not generally be cause for the volume of a phone or radio to be changed.

What is needed, therefore, is not a means for automatically adjusting the volume of an audio device to compensate for changes in all ambient noise, but rather only that noise of a frequency and duration that detracts from the ability to listen to the audio device. That is, the AVC should have some means of discriminating significant noise, which persistently detracts from listening ability, from noise that is less consequential. One means of identifying such significant noise is to measure its interference with the intelligibility of speech. One measure of interference with intelligibility considered suitable for field use is the preferred speech interference level (PSIL), which is the arithmetic average of the noise levels in the three octave bands centered at 500, 1000, and 2000 Hz [2].

Another example of an audio device is a two-way voice communications device, such as a telephone. Mobile phones in particular are often used outdoors, in crowds, and in cars and other environments where the background noise fluctuates in intensity. To adjust the volume control constantly on a phone in a noisy environment is inconvenient and often impractical. For this reason, a user of a communications device, such as a mobile phone, could potentially benefit from an

AVC feature.

The AVC for a phone is similar to the AVC for a radio in that both should have some means of discriminating significant noise from less consequential noise. Both should also have some means of separating the significant noise from a signal that requires no compensation or different compensation. In the case of a radio, the signal that requires no compensation by an AVC is the normal audio output of the radio speakers. The AVC for a radio should have some means of separating the speaker signal from the noise background. In the case of a telephone, the signal that requires no compensation or different compensation than the noise background is the telephone user's own voice. The AVC for a telephone or other multiplexed communications device should have some means of separating the user's voice from the noise background.

## II. SUMMARY OF *SmartAVC*™

A means of identifying and separating human voice from a noise background is presented in [3]. But the simplest and most practical solution for separating the user's own voice from background noise, when using an AVC-equipped mobile phone, is to momentarily suspend operation of the AVC during each instant that the user is speaking. For example, the AVC function might be suspended whenever the sound level into the microphone exceeds a threshold value indicating that the user is speaking. Since the user generally cannot understand much of what is being said to him while he is at the same time speaking on the phone, little is lost by the temporary suspension of the AVC function. The following discussion of a means of separating an audio signal from background noise, therefore, is mostly applicable to one-way communications devices, such as car radios.

For an audio amplifier providing an audio signal to audio speakers, the *SmartAVC*™ automatic volume control compensates for speech interference noise by a means including the following components and processes, as shown in Fig. 1: a microphone for detecting the background noise and the audio signal either from the speakers of a radio or the phone user's voice, and in response for producing a corresponding signal; a phase correlator process for phase correlating the microphone and audio signals; an amplitude correlator for correlating the phase-correlated microphone and audio signals; a subtraction process for producing a signal corresponding to a difference between the phase- and amplitude-correlated microphone and audio signals; a transform process for producing over a period of time a signal corresponding to the amplitude of each frequency component of the noise background; a bandpass filter for filtering the transform-produced signal to pass only frequency components within selected bands; a speech-interference level (SIL) calculation process for producing a signal corresponding to a combination of the amplitudes of the bandpass-filtered frequency components; and a solver process for producing according to an algorithm a signal for controlling the gain of the audio amplifier. Preferably the selected bands include the three octave bands centered at 500, 1000 and 2000 Hz. Preferably the transform process comprises a fast Fourier transform module. Preferably the combination of the amplitudes of the bandpass-filtered frequency components is an arithmetic average of the noise levels in the octave bands. Preferably some or all processes, algorithms, and filtering are performed by a digital signal processor (DSP) that receives both the digitized microphone signal and audio signal.

## III. SUMMARY OF CONVENTIONAL AVC

The first modern digital AVC was described in [4]. All later variations of AVCs, including *SmartAVC*™, differ from the AVC in [4] primarily by the components and processes within the digital signal processor (DSP). Other than *SmartAVC*™, the methods for controlling volume largely depend on maintaining constant signal-to-noise ratio, in some manner or other, as is done for example in [5], which keeps constant the ratio of signal to A-weighted noise.

Fig. 2 shows the main components of a conventional audio device having a conventional AVC. In Fig. 2, the components of the conventional audio device preceding its amplifier stage are not shown individually, but are generally represented by a function entitled "Signal Source." In a conventional audio device, the signal source 3 provides an electrical signal that is amplified by an audio amplifier for driving a set of speakers. The speakers convert the amplified signal to an acoustic signal that can be transmitted to listeners. Generally, the volume of such a conventional audio device is controlled by a manual volume control that adjusts the gain of the audio amplifier. The microphone receives both the transmitted signal from the speakers and any background noise. The microphone transduces the incident acoustic waves to a corresponding analog electrical signal that is communicated to an analog-to-digital (A/D) converter, wherein the analog signal is converted to a corresponding digital signal that is communicated to the DSP for processing. Concurrently, the amplified electrical signal from the audio amplifier is converted by an A/D converter to a corresponding digital signal that is also communicated to the DSP. After comparing the signals from the microphone and the audio amplifier, the DSP automatically performs a process that results in a control signal that is communicated to the audio amplifier to adjust the gain of the amplifier and, thereby, the volume of the speakers.

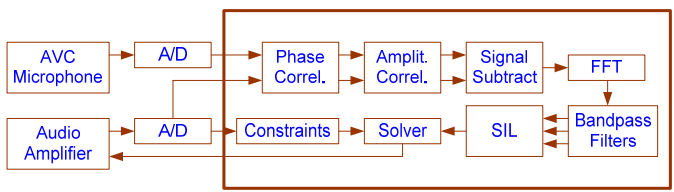

Fig. 1. Functional block diagram of DSP (within thick-lined block) of *SmartAVC*™, and DSP's interfaces with the rest of AVC and amplifier.

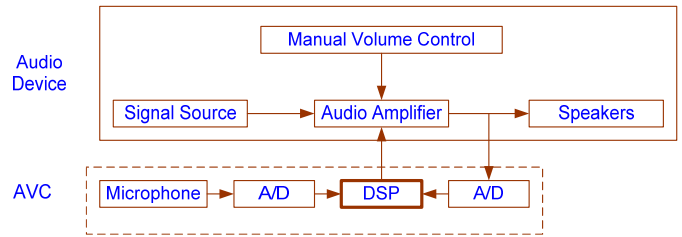

Fig. 2. Functional block diagram of conventional AVC and interface with audio device..

IV. *SmartAVC™ Preferred Embodiment*

*SmartAVC™* incorporates a novel DSP that includes the components and processes shown in Fig. 1. The correlators and the signal subtraction process cooperate to separate the sound of the speakers from the background noise so that the background noise can be processed separately. The correlators correlate the digitized inputs from the two A/Ds, so that they can be subtracted from each other by the signal subtraction process with the remainder being the background noise.

It might be possible, using factory settings, to subtract the inputs to the correlators directly without first correlating them, but the tolerance for jitter between the inputs to the correlators is so demanding that over time the system characteristics may drift and detune. The phase and amplitude correlators can correlate the inputs continuously in near real time, if necessary, or only at each start-up of the audio device, if such is sufficient. Both the phase and amplitude can be correlated with respect to the inputs over multiple processing periods for greater accuracy.

Referring again to Fig. 1, the phase correlator precedes the amplitude correlator. The phase correlator calculates the correlation function of the digitized inputs with respect to phase difference (over a limited range around the factory-set value of zero), and adjusts the relative phase to the maximum of the correlation function. The phase-correlated signals are then sent to the amplitude correlator as inputs. The amplitude correlator calculates the correlation function with respect to the gain of the audio amplifier (over a limited range around the factory-set value of one), and adjusts the gain to the minimum of the correlation function. The phase- and amplitude-correlated signals are then sent to the signal subtraction process. The signal subtraction module subtracts them to produce a difference signal that is communicated as an input to the FFT module. The difference signal is the best representation of the pure noise background after the sound from the speakers, if any, has been subtracted.

The operating characteristics of a preferred embodiment of an FFT module, optimized for minimum throughput demand, can be best described as follows. Let the sampling rate of the A/D converters be $s$ samples/second. Let the number of samples to be processed in each processing period of the FFT module be $N$, where $N$ must be an integer-power of 2. Then each processing period is $N/s$, and the time from receiving the first sample to the last in each processing period is

$$T = (N-1)/s. \qquad (1)$$

The frequency resolution of the Fourier transform is

$$\Delta f = 1/T = s/(N-1). \qquad (2)$$

The highest frequency component of the Fourier transform is

$$f_m = N\Delta f/2 = [N/(N-1)]s/2. \qquad (3)$$

In the preferred embodiment of SmartAVC™, the FFT module described below is particularly well suited to calculating the PSIL from the noise background. The PSIL is the arithmetic average of the noise levels in the three octave bands centered at 500, 1000, and 2000 Hz, that is, the three octave bands from 354 to 707 Hz, from 707 to 1414 Hz, and from 1414 to 2828 Hz, respectively.

The following design guidelines are preferred for an accurate calculation of the PSIL:

(a) The frequency resolution of the Fourier transform should be finer than about 40 Hz, that is,

$$\Delta f = s/(N-1) \leq 40 Hz, \qquad (4)$$

in order to get good statistics on the noise level by having at least of the order of 10 frequency components, even in the lowest octave band.

(b) The processing period of the FFT module should be no longer than about 25 ms, that is,

$$T = (N-1)/s \leq 25 \text{ ms}, \qquad (5)$$

in order to provide at least of the order of 10 PSIL calculations to the solver every quarter second or so. A quarter second is less than or about the time over which the AVC should begin to respond to a rapidly changing noise background.

(c) The highest frequency component of the Fourier transform should be at least about 2800 Hz, that is,

$$f_m = [N/(N-1)]s/2 \geq 2800 Hz, \qquad (6)$$

in order to get good statistics on the noise level in the highest octave band by populating it fully.

Combining these design guidelines, Eqs. (4) – (6), leads to the following point design as an example of an FFT module that is particularly well suited to calculating the PSIL for an AVC: $N = 128$; $s = 5600$ Hz; $T = 22.7$ ms; $\Delta f = 44.1$ Hz; $f_m = 2822$ Hz.

After each processing period, the FFT module sends a signal as an input to the bandpass filters, the signal comprising an amplitude for each of the frequency components of the FFT spectrum. With the point design in the preferred embodiment, the FFT calculates 65 amplitudes each processing period for the frequency components $f_j = j\Delta f = j(44.1 \text{ Hz})$, where $j = 0, 1, 2, ..., 64$. In the preferred embodiment, the frequency components, $f_9 = 397$ Hz through $f_{16} = 706$ Hz, populate the lowest octave of the PSIL. The 16 frequency components, $f_{17} = 750$ Hz through $f_{32} = 1411$ Hz, populate the middle octave of the PSIL. The 32 frequency components, $f_{33} = 1455$ Hz through $f_{64} = 2822$ Hz, populate the highest octave of the PSIL.

The bandpass filters pass only those frequency components within bands that are used by the SIL calculator. In the preferred embodiment, the bands include the 56 frequency components from $f_9$ through $f_{64}$. The SIL calculator calculates the arithmetic average (in dB) of the noise levels in the three (octave) frequency bands passed by the filters and sends as an input to the solver a single PSIL value (in dB) every processing period ($N/s = 22.9$ ms in the preferred embodiment).

The solver calculates a gain control signal, subject to certain constraints to be sent to the audio amplifier every processing period. The purpose of the solver is to calculate a gain control signal that responds proportionately to changing noise levels of a duration sufficient to interfere with intelligibility of speech or appreciation of music, and that responds negligibly to fluctuations of noise levels at the processing cycle frequency, $s/N$, or to brief noise transients. The response of the gain control signal must be somewhat dilatory to allow the solver to distinguish SIL changes of significant duration from insignificant transients. But it should not be so dilatory as to seem to the listener to be unresponsive to substantial changes of SIL.

In the preferred embodiment, the model used for the solver is that of a driven damped harmonic oscillator. The gain control signal (in dB), $a(t)$, as a function of time $t$ satisfies the second-order differential equation,

$$a''(t) + b\omega_0 a'(t) + \omega_0^2 a(t) = \omega_0^2 [S(t) + R_0], \quad (7)$$

where a prime denotes a derivative with respect to time, $b$ is a damping constant, $\omega_0$ is a constant frequency indicative of the 'stiffness' of the response, $S(t)$ is the SIL (in dB), and $R_0$ is the listener's preferred signal-to-SIL ratio (in dB). ($R_0$ is one of the constraints imposed on the solver by user interaction through the manual volume control.)

In terms of a normalized gain control signal, $A(t) \equiv a(t) - R_0$, Eq. (7) may be written as

$$A''(t) + b\omega_0 A'(t) + \omega_0^2 [A(t) - S(t)] = 0. \quad (8)$$

For the $i$th processing cycle, this model is implemented in the solver by the following algorithm:

$$A'_{i+1} = A'_i + (N/s)A''_i; \quad (9a)$$

$$\text{if } |A_i - S_i| \geq r_0, \text{ then } A_{i+1} = A_i + (N/s)A'_i; \quad (9b)$$

$$\text{otherwise } A_{i+1} = A_i; \quad (9c)$$

$$A''_{i+1} = \omega_0^2 S_{i+1} - b\omega_0 A'_{i+1} - \omega_0^2 A_{i+1}; \quad (9d)$$

$$\text{if } A_{i+1} \leq A_{\min}, \text{ then } A_{i+1} = A_{\min}. \quad (9e)$$

The constant $r_0$ (in dB) is a threshold difference of the normalized gain control signal, $A(t)$, from the SIL, $S(t)$, below which the gain control signal remains unchanged. The constant $A_{\min}$ (in dB) is the user-preferred floor of the normalized gain control signal, $A(t)$.

The constant $r_0$ is intended to desensitize the algorithm to most of the high-frequency fluctuations of the SIL in an otherwise constant noise background, and to keep $A(t)$ constant in such an environment. A typical factory setting for $r_0$ might be about 1 dB. The constant $r_0$ could also be made adaptive by making it proportional to the root-mean-square fluctuation of the SIL, for example, at the cost of additional processing.

The constant $A_{\min}$ is the listener's preferred minimum normalized gain control signal, which is generally independent of how quiet the environment may become. The listener establishes or re-establishes $A_{\min}$ through the manual volume control by adjusting the volume higher in quiet environments.

The initial conditions for the algorithm in Eqs. (9) at system start-up ($t = 0$), or whenever the user establishes new constraints through the manual volume control, are: $A_0 = S_0$, $A'_0 = 0$, $A''_0 = 0$.

Fig. 3 shows the result of implementing the algorithm of Eqs. (9) on a simulated SIL. SIL noise was simulated in Fig. 3 with significant changes of various durations and with random high-frequency fluctuations up to ±1 dB. The simulated SIL includes two transient triangular noise spikes, each 100 times (20 dB) louder than the background. For this simulation, the processing period, $N/s$, was taken to be 22.7 ms, as in the example above. The following values of constants were used in implementing the algorithm, Eqs. (9), in Fig. 3: $\omega_0 = 8\,\text{s}^{-1}$, $b = 4$, $r_0 = 1\,\text{dB}$, $A_{\min} = 2.5\,\text{dB}$. Fig. 3 also shows that the algorithm, Eqs. (9), for the normalized gain control signal, the solid black curve, responds as desired to the SIL. After a brief delay, $A(t)$ responds fully to long-duration changes in the SIL. $A(t)$ is virtually oblivious to high-frequency fluctuations.

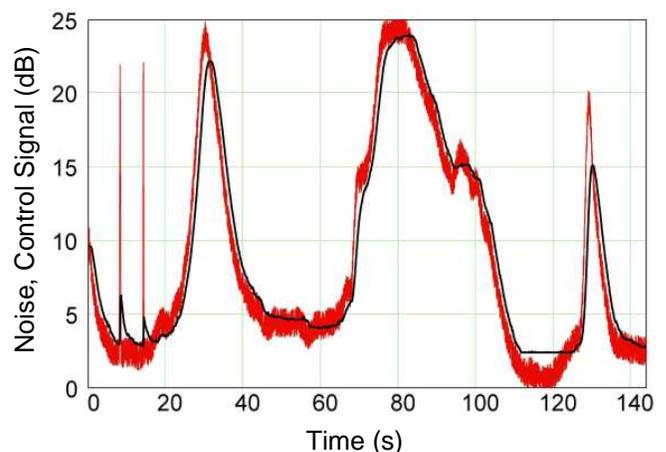

Fig. 3. Simulated SIL (red) vs. time and corresponding normalized gain control signal $A$ (black) produced by *SmartAVC™* from Eqs. (9).

To the half-second noise spike at $t = 8$ s and the quarter-second noise spike at $t = 14$ s, both 100 times louder than the background, the response of $A(t)$ is a few dB for no more than about one second. Lastly, the normalized gain control signal does not fall below the user-preferred floor of $A_{\min} = 2.5$ dB.

## V. Human Factors Features

To be fully automatic, an AVC should impose no need for additional manual controls on an audio device, other than possibly an on-off switch for the AVC feature. Listener preferences for volume should be established through normal operation of the audio device and a minimum of manual volume adjustments. The two key listener preferences that should be automatically registered by an AVC are the preferred signal-to-noise ratio and the preferred signal floor. The relevant signal-to-noise ratio is the ratio of the amplifier gain of an audio device to a suitable measure of significant noise, such as the PSIL. The preferred signal floor is the lowest amplifier gain acceptable to the listener, independent of how quiet the environment may be.

For human factors considerations, constraints are applied as inputs to the solver. Generally, it is preferable to apply at least two constraints: (1) $R_0$, the listener's preferred signal-to-SIL ratio (in dB); and (2) $A_{\min}$, the listener's preferred floor for the normalized gain control signal (in dB). There are many variations of algorithms for providing these and other constraints from the constraint module. One example follows.

Any time the manual volume control is adjusted (including at start-up of the audio device in Fig. 2), a new value of $R_0$ is calculated and sent as an input to the solver. The new value of $R_0$ is the difference between the gain control signal $a(t)$ at the end of each manual volume adjustment (or at start-up) and some weighted average of SILs calculated for the same time. For example, let the processing period during which the manual adjustment ends be denoted by the subscript $m$, and let the weighted average be over $m$ processing periods. An example of an algorithm for calculating $R_0$ is

$$R_0 = a(t_m) - \frac{1}{m}\sum_{i=1}^{m} w_i SIL_i, \qquad (10)$$

where $w_i$ is a normalized weighting function. An example of a normalized weighting function that weights SILs in processing periods near the end of an adjustment more heavily is $w_i = 2i/(m+1)$. A typical time for calculating a weighted average of SILs might be about a quarter second, or about 11 processing periods in the example given above.

Any time a weighted average of SILs is below some threshold value $SIL_t$, and the manual volume control is adjusted upward, a new value of $A_{\min}$ is calculated and sent as an input to the solver. (The threshold $SIL_t$ may be, for example, the lowest weighted average of SILs since start-up that did not prompt a manual volume adjustment during some latency period.) The new value of $A_{\min}$ is the normalized gain control signal established manually by the end of each such adjustment. When these conditions are met for establishing a new $A_{\min}$, a new $R_0$ is not also calculated. That is, if $A_{\min}$ is changed by a manual volume adjustment, $R_0$ remains unchanged by that adjustment. Any further manual volume adjustments establish new values of $A_{\min}$ and $R_0$, in accordance with the same algorithms.

Some two-way communications devices, such as mobile phones, may be equipped with two microphones, a voice microphone that selectively transduces a user's voice and a noise microphone that non-selectively transduces ambient sounds. The term "selective voice microphone" refers to a unidirectional microphone that selectively receives a voice signal from a relatively narrow solid angle in the direction of the user's voice, and that generally has low gain. A selective voice microphone is generally designed to capture the voice signal of a user, and reject most of the background noise from directions other than that of the user's voice. The term "non-selective noise microphone" refers to a microphone that is more nearly omni-directional, and that generally has higher gain, for detecting all ambient sounds, such as voices in a conference room. When an audio device is equipped with two microphones, the non-selective noise microphone will generally characterize the noise background more accurately.

## VI. Conclusions

This paper presented a method that automatically controls volume to preserve constant intelligibility of speech, rather than constant signal-to-noise ratio, and thereby avoids overcompensating for noises in the frequency domain that do not degrade intelligibility. Additionally, the method uses an algorithm resembling that of a shock absorber to smooth out fluctuations of noise in the time domain that do not affect intelligibility. The patented *SmartAVC™* algorithms have been demonstrated on an A-B breadboard unit to provide substantial advantages in performance [6] by controlling volume without sacrificing clarity, not overcompensating for high-frequency noise, smoothing volume fluctuations, and eliminating static-like bursts caused by noise spikes.